\documentclass{article}
\usepackage{spconf,amsmath,graphicx,hyperref}
\usepackage{graphicx}
\usepackage{booktabs}
\usepackage{hyperref}
\usepackage{multirow}
\usepackage{makecell}
\usepackage{amsmath, amssymb}
\usepackage{float}
\usepackage[font=small,labelfont=bf,labelsep=colon]{caption}
\usepackage{subcaption}

\title{Dual-Strategy-Enhanced ConBiMamba for Neural Speaker Diarization}
\name{Zhen Liao, Gaole Dai, Mengqiao Chen, Wenqing Cheng, Wei Xu*\thanks{*Wei Xu is the corresponding author.\newline \hspace*{1.5em} This work is supported by the National Key Research and Development Program of China under Grant 2021YFC3340803.}}
\address{School of Electronic Information and Communications\\
Hubei Provincial Key Laboratory of Smart Internet Technology\\
Huazhong University of Science and Technology, China\\
\small \texttt{\{zh3n,daigle,chenmengqiao,chengwq,xuwei\}@hust.edu.cn}}
\begin{document}
\ninept
\maketitle
\begin{abstract}
Conformer and Mamba have achieved strong performance in speech modeling but face limitations in speaker diarization. Mamba is efficient but struggles with local details and nonlinear patterns. Conformer’s self-attention incurs high memory overhead for long speech sequences and may cause instability in long-range dependency modeling. These limitations are critical for diarization, which requires both precise modeling of local variations and robust speaker consistency over extended spans. To address these challenges, we first apply ConBiMamba for speaker diarization. We follow the Pyannote pipeline and propose the Dual-Strategy-Enhanced ConBiMamba neural speaker diarization system. ConBiMamba integrates the strengths of Conformer and Mamba, where Conformer’s convolutional and feed-forward structures are utilized to improve local feature extraction. By replacing Conformer’s self-attention with ExtBiMamba, ConBiMamba efficiently handles long audio sequences while alleviating the high memory cost of self-attention. Furthermore, to address the problem of the higher DER around speaker change points, we introduce the Boundary-Enhanced Transition Loss to enhance the detection of speaker change points. We also propose Layer-wise Feature Aggregation to enhance the utilization of multi-layer representations. The system is evaluated on six diarization datasets and achieves state-of-the-art performance on four of them. The source code of our study is available at \url{https://github.com/lz-hust/DSE-CBM}.
\end{abstract}
\section{Introduction}
\label{sec:intro}
Speaker diarization is the task of determining ``who spoke when'' in an audio recording. Clustering-based diarization systems are composed of a cascade of multiple subsystems \cite{landini2024modular}. However, these subsystems are often optimized independently, leading to performance bottlenecks. Moreover, once an audio recording is segmented, each segment is represented by a speaker embedding, and a clustering algorithm merges segments based on embedding similarity. This design assumes that each segment contains speech from a single speaker, rendering such systems incapable of handling overlapping speech. To overcome these limitations, Fujita et al. proposed End-to-End Neural Diarization (EEND) \cite{fujita2019eend,saeend,sceend,eendeda}, which formulates diarization as a multi-label classification problem and thereby supports overlapping speech modeling. Nevertheless, the computational complexity of EEND increases rapidly with the number of speakers and the length of the audio. Based on this, End-to-End Neural Diarization with Vector Clustering (EEND-VC) \cite{eendvc1,eendvc2} introduces a clustering mechanism. It operates by segmenting the audio into chunks, processing them with a local EEND, employing an encoder to extract embeddings from active regions, and clustering them to stitch a global diarization result. EEND-VC enables the modeling of variable numbers of speakers and long-duration recordings. Following the EEND-VC framework, Bredin et al. proposed Pyannote \cite{bredin2023pyannote,powerset}, which achieved excellent performance on multiple datasets.

In the speaker diarization task, models are required to achieve precise characterization of local details and robust modeling of long-term dependencies. Pyannote, which builds its local EEND module on BiLSTM \cite{lstm}, has shown strong performance across multiple datasets. However, BiLSTM relies on sequential recursive computation, making it insufficient for capturing long-range dependencies. Han et al. \cite{diarizen} proposed replacing the BiLSTM with Conformer \cite{conformer} in the local EEND, but Conformer’s self-attention \cite{attention} incurs high computational and memory costs on long sequences, leading to instability when modeling long-range dependencies—issues that are particularly problematic in long-duration sessions. The selective State Space Model Mamba \cite{mamba} may provide a feasible direction to mitigate the aforementioned limitations. Mamba efficiently captures long-range dependencies with computational complexity that scales linearly with sequence length. However, Mamba primarily emphasizes global modeling, often at the expense of local detail representation. This limitation is particularly critical for diarization, where accurate modeling of speaker change points is essential. Moreover, Mamba relies solely on past context for sequence modeling. To address this, Zhang et al. proposed ExtBiMamba \cite{speechmamba}, which analyzes both past and future speech segments to overcome the unidirectionality of Mamba. Nevertheless, ExtBiMamba remains largely focused on long-range information integration and lacks explicit mechanisms for local pattern modeling. To alleviate this, Zhang et al. proposed ConBiMamba by replacing the multi-head self-attention in Conformer with ExtBiMamba. ConBiMamba retains the convolution module of Conformer to enhance local feature extraction, while leveraging ExtBiMamba for efficient long-range dependency modeling. This hybrid design has demonstrated strong performance in speech enhancement and automatic speech recognition.
\begin{figure*}[htbp]
    \centering
    \begin{subfigure}[b]{0.32\textwidth}
        \centering
        \includegraphics[width=1.0\linewidth,height=6.5cm,keepaspectratio]{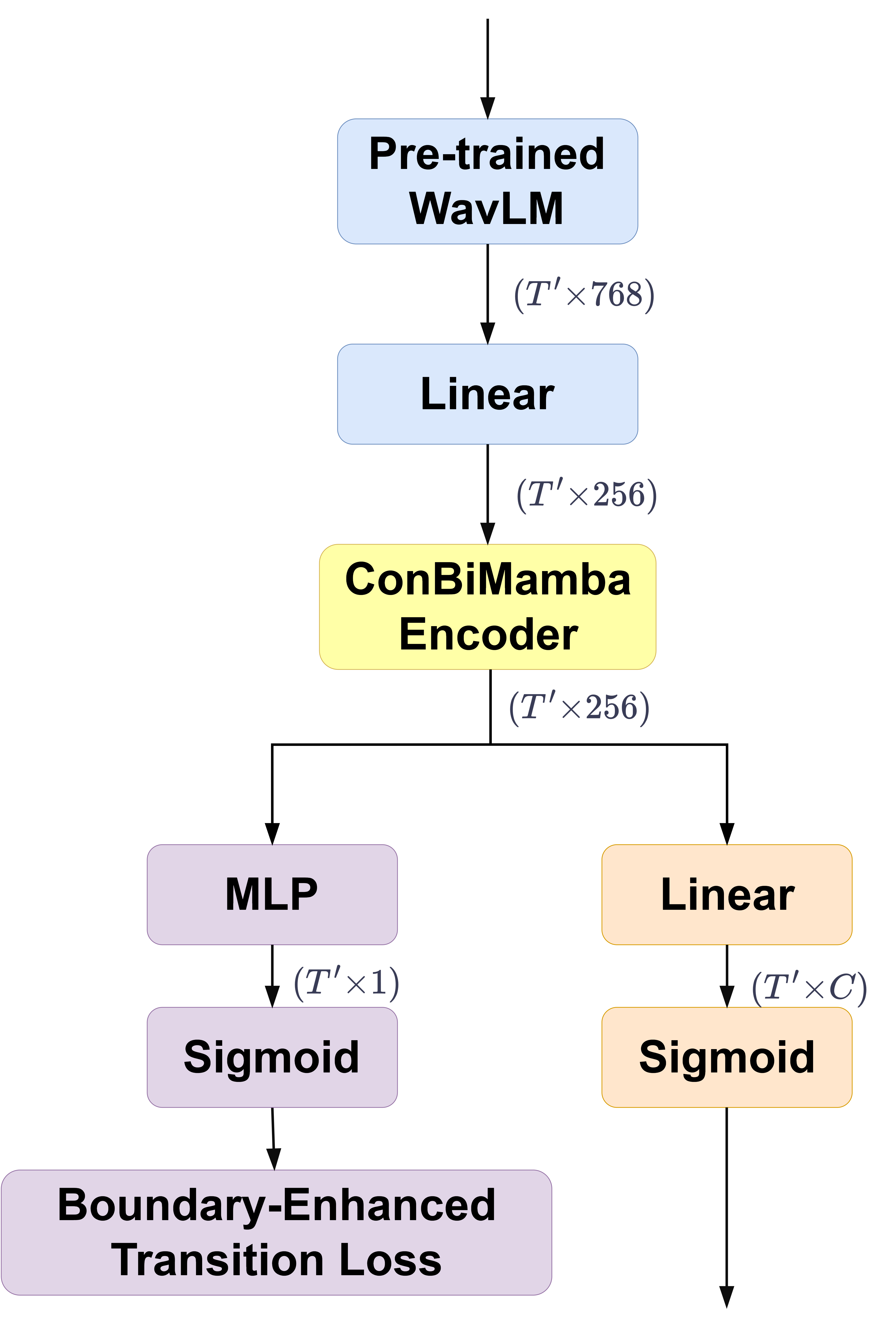}
        \caption{Local EEND based ConBiMamba}
        \label{fig:module1}
    \end{subfigure}\hspace{0mm}
    \begin{minipage}[b]{0.36\textwidth}
        \centering
        \begin{subfigure}{\linewidth}
            \centering
            \includegraphics[width=1.0\linewidth,height=4.5cm,keepaspectratio]{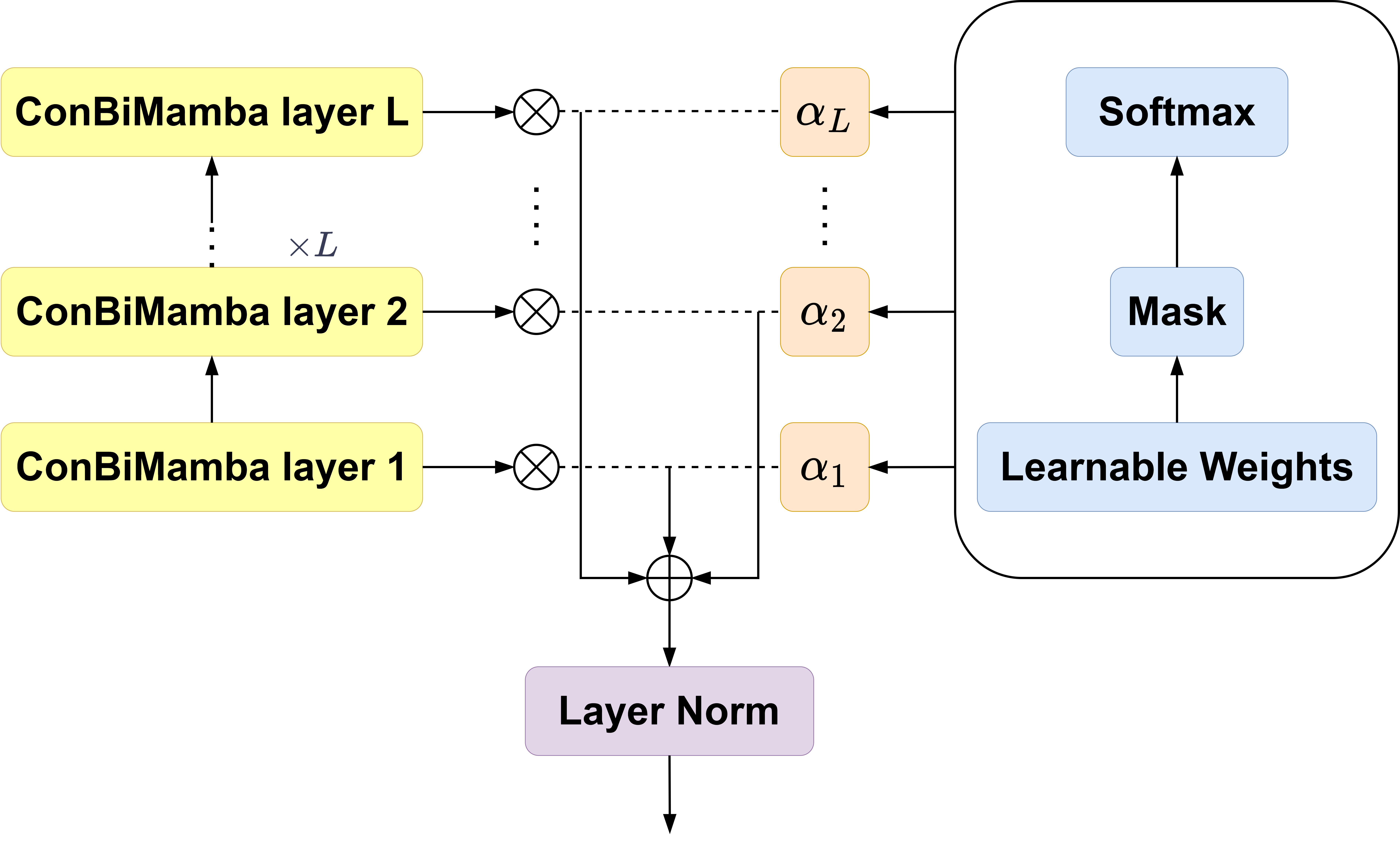}
            \caption{ConBiMamba Encoder}
            \label{fig:module2}
        \end{subfigure}
        \vspace{0mm}
        \begin{subfigure}{\linewidth}
            \centering
            \includegraphics[width=1.0\linewidth,height=2cm,keepaspectratio]{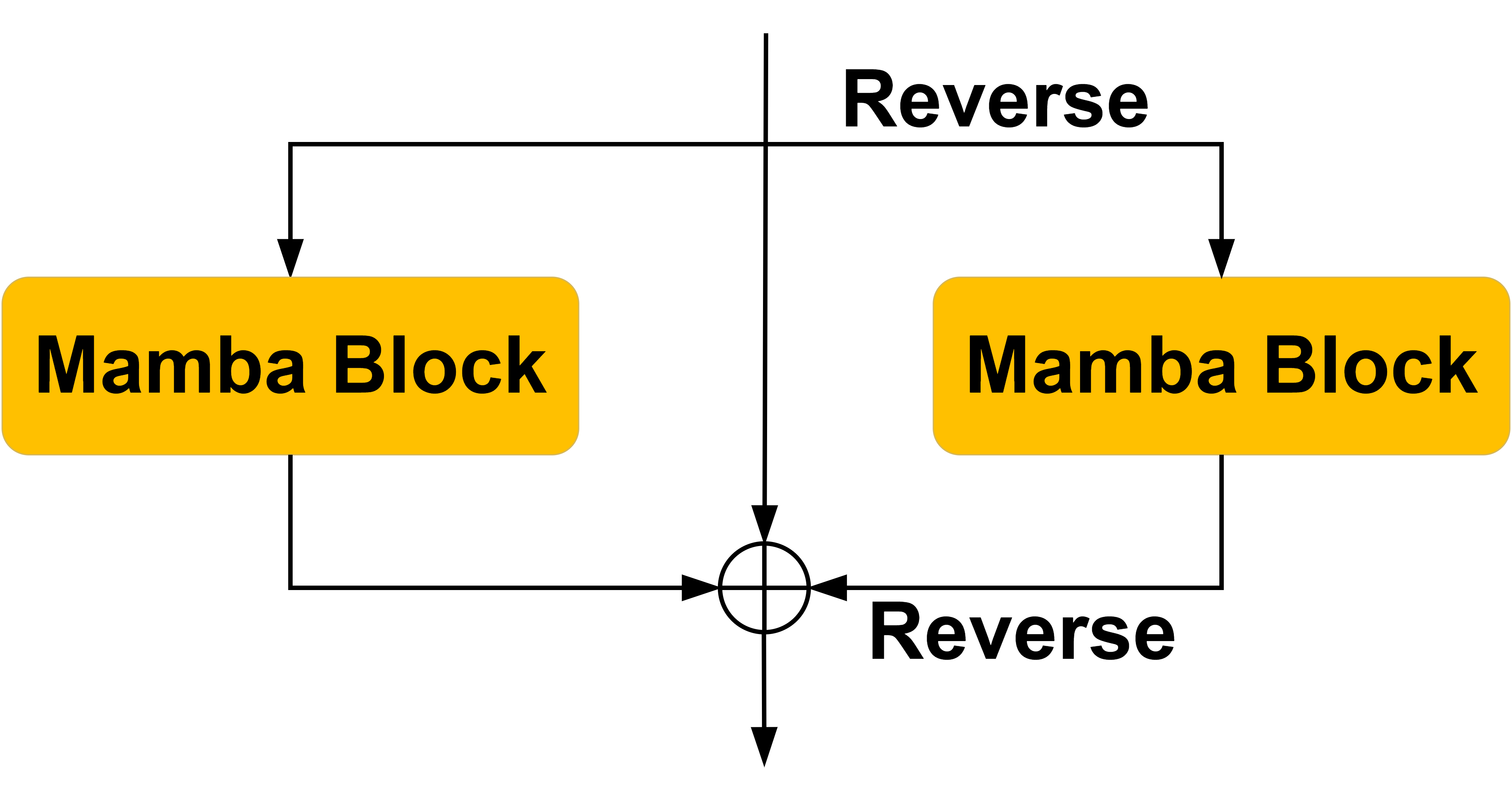}
            \caption{ExtBiMamba}
            \label{fig:module4}
        \end{subfigure}
    \end{minipage}\hspace{0mm}
    \begin{subfigure}[b]{0.24\textwidth}
        \centering
        \includegraphics[width=1.0\linewidth,height=6.5cm,keepaspectratio]{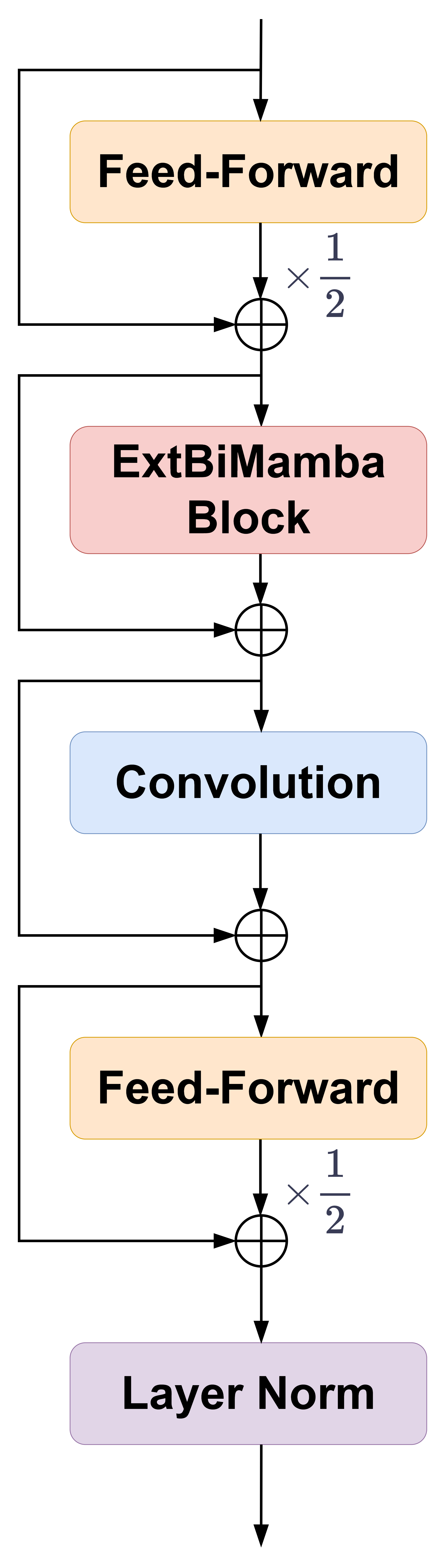}
        \caption{ConBiMamba layer}
        \label{fig:module3}
    \end{subfigure}
    \caption{Overall framework of the ConBiMamba-based local EEND.}
    \label{fig:four_modules}
\end{figure*}
In this work, we apply ConBiMamba into the speaker diarization task, and develop the Dual-Strategy-Enhanced ConBiMamba Neural Speaker Diarization system. We propose two strategies. First, we introduce an auxiliary task called speaker change point detection, which captures speaker state changes between adjacent frames. This addresses the high diarization error rate (DER) near speaker change points \cite{knox2012did}, as existing methods typically rely on frame-level supervision, which requires models to simultaneously perform speaker identification and boundary localization and often leads to unstable predictions at speaker change points. Second, to address the inefficiency of most methods that utilize only the last-layer output, neglecting the complementarity of intermediate and deep features, we propose the mask-based Layer-wise Feature Aggregation to integrate representations from different layers, thereby enhancing the model’s representation capability.
\section{METHODS}
\label{sec:pagestyle}
\subsection{Local EEND based ConBiMamba}
In view of the complementarity of Conformer and Mamba in speech modeling and their deficiencies in diarization, to simultaneously take into account local detail modeling and long-range dependency capture, we follow the Pyannote pipeline and apply ConBiMamba to local EEND, with the overall architecture illustrated in Fig.~\ref{fig:four_modules}(\subref{fig:module1}). We use a frozen, pre-trained WavLM Base+ model as the feature extractor, producing 768-dimensional features per frame \cite{wavlm}. These features are projected to 256 dimensions via a linear layer and fed into a ConBiMamba Encoder (structure shown in Fig.~\ref{fig:four_modules}(\subref{fig:module2})) composed of seven stacked ConBiMamba Layers. The Mamba block hyperparameters in each ConBiMamba Layer (shown in Fig.~\ref{fig:four_modules}(\subref{fig:module3})) follow the configurations in \cite{speechmamba}. The original convolution module employed a single depthwise convolution kernel of size 31. To enhance the model’s multi-scale perception of temporal data, we redesign it as a multi-branch structure with convolutional kernels of sizes $K = \{15, 31, 63\}$, where each branch performs independent depthwise convolution and the outputs are averaged. The primary speaker diarization task produces its final output by passing the ConBiMamba Encoder representations through a linear layer followed by a Sigmoid function. This process represents multi-class states. In all experiments, we assume a maximum of 4 speakers and 2 overlapping speakers per audio segment. Inspired by \cite{mamba-diarization}, we do not adopt Powerset representation \cite{powerset}, as it tends to increase the DER in practical pipelines. In addition, for the auxiliary task of speaker change detection, we design an independent branch in which the ConBiMamba Encoder outputs are fed into a multi-layer perceptron and then processed by a Sigmoid function to produce per-frame speaker change signals. Details of this branch are provided in Section~\ref{speaker change}.
\subsection{Layer-wise Feature Aggregation}
Most existing methods rely solely on the encoder’s last-layer output, neglecting the complementarity of features across layers. Our Layer-wise Feature Aggregation strategy aims to integrate complementary information by adaptively aggregating the output features from different ConBiMamba Layers, thereby enhancing the model’s robustness to complex sequences. However, we argue that shallow features, while useful for low-level modeling, are less relevant to speaker diarization’s discriminative objective and may introduce redundancy and noise. Theoretically, the learnable weights could suppress low-relevance features, but optimization challenges, gradient bias, and multi-task interference often prevent this. Thus, we aggregate only the final few layers’ outputs to retain deep representations and incorporate complementary mid-level information, enhancing the model's representation capability.

We first introduce a set of learnable scalar parameters 
$\boldsymbol{\alpha} = [\alpha_1, \alpha_2, \dots, \alpha_L] \in \mathbb{R}^L$, ($L = 7$), representing the initial importance weights of each layer's features. These initial weights are then modulated by a predefined static mask 
$\mathbf{m} = [m_1, m_2, \dots, m_L] \in \mathbb{R}^L$:
\begin{equation}
\tilde{\boldsymbol{\alpha}} = \boldsymbol{\alpha} \odot \mathbf{m} + (1 - \mathbf{m}) \cdot (-\infty),
\end{equation}
where $\odot$ denotes element-wise multiplication, and $-\infty$ is used to suppress the weights of masked layers (where $m_l = 0$), ensuring their contribution is effectively zero and enabling a layer selection mechanism. The adjusted weights are then normalized using the Softmax function to obtain the relative contribution of each layer:
\begin{equation}
\mathbf{w} = 
\left[
\frac{e^{\tilde{\alpha}_1}}{\sum_{k=1}^L e^{\tilde{\alpha}_k}}, \ 
\frac{e^{\tilde{\alpha}_2}}{\sum_{k=1}^L e^{\tilde{\alpha}_k}}, \ \dots, \
\frac{e^{\tilde{\alpha}_L}}{\sum_{k=1}^L e^{\tilde{\alpha}_k}}
\right] \in \mathbb{R}^L.
\end{equation}

Finally, aggregated features are computed as the weighted sum of the layer outputs, followed by layer normalization and dropout.
\begin{equation}
\mathbf{h}_{\text{output}} = \text{Dropout}\Big(\text{LayerNorm}\Big(\sum_{l=1}^L w_l \cdot \mathbf{h}_{\text{ConBiMamba}}^{(l)}\Big)\Big).
\label{eq:layerwise_output}
\end{equation}
\subsection{Boundary-Enhanced Transition Loss}
\label{speaker change}
In speaker diarization, accurately detecting speaker change points is critical for reducing DER \cite{knox2012did}. However, existing methods typically rely solely on frame-level speaker label supervision, requiring models to implicitly learn both speaker identification and boundary localization, which often leads to unstable predictions at boundaries. To address this, we introduce an auxiliary task, speaker change point detection, to explicitly model speaker state transitions between adjacent frames, enhancing the model’s sensitivity and robustness to speaker changes. This task is implemented through a dedicated Boundary-Enhanced Transition Loss.

Given the primary task label matrix $\mathbf{Y} \in \mathbb{R}^{B \times T \times K}$, where $B$ denotes the batch size, $T$ the number of frames, and $K$ the maximum number of speakers, each element $y_{b,t,k} \in \{0,1\}$ ( $t = 1, \dots, T-1$) indicates whether the $k$-th speaker is active at frame $t$ in sample $b$. The change point label $\mathbf{C} \in \mathbb{R}^{B \times (T-1)}$, is defined as:
\begin{equation}
c_{b,t} =
\begin{cases} 
1, & \exists k \in \{1, \dots, K\} \ \text{s.t. } y_{b,t+1,k} \neq y_{b,t,k}, \\
0, & \text{otherwise},
\end{cases}
\label{eq:change_label}
\end{equation}
This label marks a change point whenever any difference in speaker state is observed between adjacent frames.

The output features obtained from Eq.~\eqref{eq:layerwise_output} are projected from 256 dimensions to 128 dimensions, followed by a ReLU activation, and then projected to 1 dimension to produce the raw logit for each frame:
\begin{equation}
o_{b,t} = W_2 \cdot \text{ReLU}(W_1 \cdot h_{b,t} + b_1) + b_2.
\label{eq:change_logit}
\end{equation}

To dynamically adjust the loss weight, we compute the positive sample ratio. Let $N = B \times (T-1)$ denote the total number of samples, and $P = \sum_{b,t} c_{b,t}$ the number of positive samples. The positive sample ratio is then defined as $r = P / N$ (with a default value of $0.1$ if $r = 0$). To handle label imbalance, the Boundary-Enhanced Transition Loss adopts the Focal Loss formulation:
\begin{equation}
p_{b,t} =
\begin{cases} 
\sigma(o_{b,t}), & \text{if } c_{b,t} = 1, \\
1 - \sigma(o_{b,t}), & \text{if } c_{b,t} = 0,
\end{cases}
\label{eq:pb}
\end{equation}
\begin{equation}
L_{\text{BET}} = \frac{1}{N} \sum_{b=1}^B \sum_{t=1}^{T-1} 
-\alpha (1 - p_{b,t})^\gamma \log(p_{b,t}),
\label{eq:change_loss}
\end{equation}
where $\sigma(\cdot)$ denotes the Sigmoid function, $\alpha$ controls the contribution of positive samples, and $\gamma$ down-weights easy samples while emphasizing harder ones. The total loss combines the permutation-invariant training loss $L_{\text{PIT}}$ \cite{fujita2019eend} and the Boundary-Enhanced Transition Loss, weighted by $\lambda$:
\begin{equation}
L_{\text{total}} = L_{\text{PIT}} + \lambda L_{\text{BET}}.
\label{eq:total_loss}
\end{equation}
In all experiments, we set $\alpha = r$, $\gamma = 2$, and $\lambda = 0.5$.

\section{EXPERIMENTS}
\label{sec:pagestyle}
\subsection{Datasets}
We use six widely used datasets: AISHELL-4 \cite{aishell}, MagicData-RAMC \cite{ramc}, VoxConverse \cite{vox}, MSDWild \cite{msdwild}, AMI (channel 1) \cite{ami,ami2}, and AliMeeting \cite{ali}. In addition, following the approach introduced in EEND \cite{fujita2019eend,saeend,sceend,eendeda}, we constructed a simulated dataset comprising four-speaker conversations. This dataset, generated from LibriSpeech \cite{librispeech}, has a total duration of 786 hours with a sampling rate of 16 kHz, and includes MUSAN background noise \cite{musan} and room impulse responses \cite{ko2017study}. Following the method described in  \cite{powerset}, we merged the training and validation sets of this simulated dataset with those of the six datasets to form a compound dataset.
\subsection{Training setup}
Training proceeds in two stages. In the first stage, we use the compound dataset, where recordings are segmented into 20-second segments with a stride of 20 seconds. Training is run for up to 60 epochs with a batch size of 16, using the C-AdamW optimizer \cite{cadamw}. The learning rate undergoes a 5-epoch linear warm-up to 2e-4, and is subsequently halved if the validation loss does not improve for 2 consecutive epochs, down to 1e-6. Early stopping occurs after 10 consecutive epochs without improvement. In the second stage, the final model from the first stage is fine-tuned on each dataset. Recordings are segmented into 10, 20, or 30-second segments with a stride matching the segment length, and the optimal length is selected for each dataset. Training runs for up to 20 epochs with a 2-epoch warm-up, an initial learning rate of 2e-5. The learning rate is halved if the validation loss does not improve for 2 consecutive epochs, and early stopping occurs after 5 consecutive epochs without improvement.
\subsection{Inference setup}
\label{infer}
In the inference stage, we follow the Pyannote pipeline. For each experiment, the model parameters are obtained by averaging the weights of the last three checkpoints. Local speaker embeddings are extracted using the ECAPA-TDNN \cite{ecapa} model (from the SpeechBrain toolkit \cite{ravanelli2021speechbrain}). Subsequently, agglomerative hierarchical clustering \cite{ahc} with centroid linkage is applied to group the speaker embeddings. During clustering, the hyperparameters, namely the clustering threshold and the minimum cluster size, are optimized via Bayesian Optimization with Sequential Model-Based Optimization using the scikit-optimize \footnote{\url{https://scikit-optimize.readthedocs.io}}, with 50 iterations per search.
\begin{table*}[!t]
  \centering
  \caption{DER(\%) performance comparison of different systems based on the Pyannote pipeline (collar = 0)}
  \label{tab:table1}
  \begin{tabular}{c c c c c c c}
    \hline
    System & AISHELL-4 & RAMC & \makecell{VoxConverse \\ \textit{v0.3}} &\makecell{MSDWild \\ \textit{Few}} & \makecell{AMI \\ \textit{Channel 1}} & \makecell{AliMeeting \\ \textit{far}} \\
    \hline
    PyannoteAI \cite{powerset} & 11.9 & 18.4 & 9.4 & 19.8 & 20.9 & 22.5 \\

    Diarizen(WavLM-frozen) \cite{diarizen} & 11.7 & - & - & - & 17.0 & 19.9 \\
    
    Diarizen(WavLM-updated) \cite{diarizen} & 11.7 & - & - & - & \textbf{15.4} & 17.6 \\
    
    Mamba-diarization \cite{mamba-diarization} & 10.5 & 11.0 & 9.3 & 19.8 & 18.5 & 16.2 \\
    
    Proposed & \textbf{9.8} & \textbf{10.9} & \textbf{8.6} & \textbf{19.2} & 16.7 & 14.9 \\
    
    State-of-the-art by August 2025 & 10.5 \cite{mamba-diarization} & 11.0 \cite{mamba-diarization} & 9.3 \cite{mamba-diarization} & 19.6 \cite{baroudi2024specializing} & \textbf{15.4} \cite{diarizen} & \textbf{13.2} \cite{harkonen2024eend} \\
    \hline
  \end{tabular}
\end{table*}

\begin{table*}[!t]
  \centering
  \caption{Effect of Layer-wise Feature Aggregation at different layers on DER(\%) (collar=0)}
  \label{tab:table2}
  \begin{tabular}{c c c c c c c}
    \hline
    Selected layers & AISHELL-4 & RAMC & \makecell{VoxConverse \\ \textit{v0.3}} &\makecell{MSDWild \\ \textit{Few}} & \makecell{AMI \\ \textit{Channel 1}} & \makecell{AliMeeting \\ \textit{far}} \\
    \hline
    Last 7 Layers & 10.2 & 11.5 & 8.9 & 19.9 & 17.4 & 16.4 \\
    
    Last 5 Layers & 10.0 & 11.6 & 8.8 & 19.6 & 17.0 & 15.3 \\
    
    Last 3 Layers & \textbf{9.8} & \textbf{10.9} & \textbf{8.6} & \textbf{19.2} & \textbf{16.7} & \textbf{14.9} \\
    
    Last 1 Layer & 9.9 & 11.3 & 8.9 & 19.4 & 17.3 & 15.5 \\
    \hline
  \end{tabular}
\end{table*}

\begin{table*}[!t]
  \centering
  \caption{Effect of ablating the Boundary-Enhanced Transition Loss on DER(\%) (collar=0)}
  \label{tab:table3}
  \begin{tabular}{c c c c c c c}
    \hline
      & AISHELL-4 & RAMC & \makecell{VoxConverse \\ \textit{v0.3}} &\makecell{MSDWild \\ \textit{Few}} & \makecell{AMI \\ \textit{Channel 1}} & \makecell{AliMeeting \\ \textit{far}} \\
    \hline
    Complete system (Last 3 layers) & \textbf{9.8} & \textbf{10.9} & \textbf{8.6} & \textbf{19.2} & \textbf{16.7} & \textbf{14.9} \\
    - Boundary-Enhanced Transition Loss & 9.9 & 11.0 & 9.0 & 19.5 & 17.2 & 15.9 \\
    \hline
    Complete system (Last 1 layer) & \textbf{9.9} & \textbf{11.3} & \textbf{8.9} & \textbf{19.4} & \textbf{17.3} & \textbf{15.5} \\
    - Boundary-Enhanced Transition Loss & 10.0 & 11.5 & 9.0 & 19.8 & 17.4 & 16.4 \\
    \hline
  \end{tabular}
\end{table*}
\subsection{Result and analysis}
\label{comprison}
We compared our proposed system with other existing systems based on the Pyannote pipeline on six datasets. Table~\ref{tab:table1} summarizes the DER performance with a 0-second collar, which is also used in the experiments reported in Sections~\ref{layer} and~\ref{ablation}. The results for PyannoteAI are sourced from its official code repository \footnote{\url{https://github.com/pyannote/pyannote-audio}}, while the results of Diarizen \cite{diarizen} and Mamba-diarization \cite{mamba-diarization} are cited from the publications. Our system employs Layer-wise Feature Aggregation based on the last three layers. For further details, refer to Section~\ref{layer}.

Results show that the proposed system outperforms the best Pyannote system, PyannoteAI. Additionally, it surpasses Mamba-diarization and Diarizen, both of which use the WavLM Base+ \cite{wavlm} with frozen pre-trained weights as the feature extractor. We also compiled the current SOTA results. As shown in Table~\ref{tab:table1}, our model achieves results surpassing the published SOTA on four datasets: AISHELL-4, RAMC, VoxConverse, and MSDWild. Notably, our system excels in capturing speech boundaries. For instance, on MSDWild, our system achieves a false alarm rate of 5.02\% and a miss rate of 7.53\%, compared to 4.77\% and 8.88\% for Mamba-diarization \cite{mamba-diarization}, respectively. On RAMC, our false alarm and miss rates are 2.82\% and 1.56\%, respectively, compared to 4.48\% and 2.75\% for Mamba-diarization. This aligns with our expectations: ConBiMamba excels in local detail modeling, and the Boundary-Enhanced Transition Loss further enhances boundary prediction capability.

However, when Diarizen \cite{diarizen} jointly updates the WavLM Base+ and other parameters within its system, its performance improves significantly, particularly on AMI (channel 1). This improvement arises from the joint optimization of WavLM Base+ and its Conformer-based processing module. On the one hand, gradients from the Conformer module propagate back to WavLM, fine-tuning its feature extraction to better align with the speaker discrimination requirements of the diarization task. On the other hand, the fine-tuned WavLM features provide more accurate inputs to the Conformer, thereby creating a positive feedback loop of feature optimization and module adaptation. This synergy surpasses the performance limits of a fixed pre-trained model, leading to a substantial overall system improvement. In addition, we observe that Pyannote-based systems exhibit a larger gap from SOTA on the AliMeeting. As noted in \cite{diarizen}, this may be due to the fact that each session in AliMeeting involves only 2–4 speakers, a setting with fewer speakers, which may be more amenable to purely end-to-end approaches. We further speculate that another reason is the relatively high proportion of overlapping speech in AliMeeting. Existing Pyannote-based approaches, including our method, do not explicitly model overlapping speech, which prevents them from fully reaching their performance potential on this dataset. These insights offer viable solutions for optimizing our system in the future.
\subsection{Analysis of Layer-wise Feature Aggregation}
\label{layer}
We studied the effect of using different numbers of ConBiMamba layers in the Layer-wise Feature Aggregation process. As shown in Table~\ref{tab:table2}, aggregating the last three layers consistently yields the best performance. Compared with using only the last layer as output, this strategy reduces DER across all six datasets, improving system performance. To compare with SOTA methods, the experiments in Section~\ref{comprison} are conducted based on this configuration. In contrast, aggregating the last five layers yields results comparable to using only the final layer, while aggregating the last seven layers even degrades performance on most datasets. This observation aligns with our expectation that introducing excessive shallow-layer features tends to introduce noise and redundancy.
\subsection{Ablation study on Boundary-Enhanced Transition Loss}
\label{ablation}
To evaluate the effectiveness of the proposed speaker change modeling mechanism, we ablated the MLP branch for predicting speaker change signals and Boundary-Enhanced Transition Loss, retaining only $L_{\text{PIT}}$, in both the complete system with layer-wise feature aggregation over the last three ConBiMamba layers and the system using only the final layer as output. As shown in Table~\ref{tab:table3}, in both cases, this ablation leads to varying degrees of increase in DER on all datasets, indicating that the mechanism contributes positively to model stability and overall performance. This suggests that explicitly modeling speaker change is an effective strategy and also provides potential insights for exploring other auxiliary supervision tasks such as overlap detection or acoustic event modeling.
\section{CONCLUSION}
\label{sec:typestyle}
In this study, we propose the Dual-Strategy-Enhanced ConBiMamba Neural Speaker Diarization system. We comprehensively evaluate the proposed system on six widely used public datasets and systematically compare it with other representative Pyannote pipeline-based systems as well as previously published state-of-the-art methods. The results demonstrate that our ConBiMamba-based system achieves strong performance across multiple datasets, particularly in boundary detection. Moreover, the experiments further validate the effectiveness of two proposed strategies, Boundary-Enhanced Transition Loss and Layer-wise Feature Aggregation, in reducing DER.

\bibliographystyle{IEEEbib}
\bibliography{strings,main}

\end{document}